\documentclass[preprint]{aastex}
\usepackage{rotating}
\usepackage{pxfonts}
\usepackage{booktabs,longtable}
\usepackage{graphicx}
\usepackage{subfigure}

\begin{document}

\title{The Megamaser Cosmology Project. X.  High Resolution Maps and Mass Constraint for SMBHs}

\author{W. Zhao\altaffilmark{1,2}, 
J. A. Braatz\altaffilmark{3}, 
J. J. Condon\altaffilmark{3}, 
K. Y. Lo\altaffilmark{3}, 
M. J. Reid\altaffilmark{4}, 
C. Henkel\altaffilmark{5,6},
D. W. Pesce\altaffilmark{7},
J. E. Greene\altaffilmark{8},
F. Gao\altaffilmark{3,9,10}, 
C.~Y.~Kuo\altaffilmark{11}, 
and C. M. V. Impellizzeri\altaffilmark{3,12}} 
\affil{\altaffilmark{1}Shanghai Observatory, 
80 Nandan Road, Shanghai 200030, China}  
\affil{\altaffilmark{2}Key Laboratory of Radio Astronomy, 
Chinese Academy of Sciences, 210008 Nanjing, PR China}
\affil{\altaffilmark{3}National Radio Astronomy Observatory, 
520 Edgemont Road, Charlottesville, VA 22903, USA}   
\affil{\altaffilmark{4}Harvard-Smithsonian Center for Astrophysics, 
60 Garden Street, Cambridge, MA 02138, USA}
\affil{\altaffilmark{5}Max-Planck-Institut f\"ur Radioastronomie, 
Auf dem H\"ugel 69, 53121 Bonn, Germany} 
\affil{\altaffilmark{6}King Abdulaziz University, 
P.O. Box 80203, Jeddah, Saudi Arabia}
\affil{\altaffilmark{7}Department of Astronomy, University of Virginia,
 Charlottesville, VA 22904}    
\affil{\altaffilmark{8}Department of Astrophysical Sciences, 
Princeton University, Princeton, NJ 08544, USA}   
\affil{\altaffilmark{9}Key Laboratory for Research in Galaxies and Cosmology, 
Shanghai Astronomical Observatory, Chinese Academy of Science, 
Shanghai 200030, China}
\affil{\altaffilmark{10}Graduate School of the Chinese Academy of Sciences, 
Beijing 100039, China} 
\affil{\altaffilmark{11}Department of Physics, National Sun Yat-Sen University, 
No.70, Lianhai Rd., Gushan Dist., Kaohsiung City 804, Taiwan (R.O.C.)} 
\affil{\altaffilmark{12}Joint Alma Office, Alsonso de Cordova 3107, Vitacura, Santiago, Chile}

\begin{abstract}
We present high resolution (sub-mas) VLBI maps of nuclear
H$_2$O megamasers for seven galaxies. In UGC~6093, the well-aligned 
systemic masers and high-velocity masers originate in an edge-on, flat disk and we 
determine the mass of the central SMBH to be $M_\mathrm{SMBH} = 2.58\times10^7 M_\odot$ ($\pm 7\%$).
For J1346+5228, the distribution of masers is consistent with a disk, but the faint high-velocity masers 
are only marginally detected, and we constrain the mass of the SMBH to be in the range $1.5 - 2.0\times 10^7 M_\odot$.
The origin of the masers in Mrk~1210 is less clear, as the systemic and high-velocity masers are misaligned and
show a disorganized velocity structure.  We present one possible model in which the
masers originate in a tilted, warped disk, but we do not rule out the possibility of other explanations including outflow masers.
In NGC~6926, we detect a set of redshifted masers, clustered within a pc of each other, and a single blueshifted maser
about 4.4 pc away, an offset that would be unusually large for a maser disk system.  Nevertheless, if it is a disk system, we
estimate the enclosed mass to be $M_\mathrm{SMBH} < 4.8\times10^7 M_\odot$.
For NGC~5793, we detect redshifted masers spaced about 1.4 pc from a clustered set of
blueshifted features.  The orientation of the structure supports a disk scenario as suggested by \citet{Hagiwara01}.  
We estimate the enclosed mass to be $M_\mathrm{SMBH} < 1.3\times10^7 M_\odot$.
For NGC 2824 and J0350$-$0127, the masers may be associated with pc or sub-pc scale jets or outflows.

\end{abstract}

\section{Introduction}
H$_2$O megamasers are found in high-density, warm molecular 
gas with a large water abundance in the central parsec of galaxies 
with Seyfert 2 or LINER spectra, either in {\bf the circumnuclear disks}, 
or associated with the jets and outflows triggered by these 
active galactic nuclei (AGNs). {\bf In the circumnuclear disks}, the masing 
gas is heated by X-ray irradiation or shocks and viscosity 
within the disk \citep{neu94}. In the jets and outflows impacting 
the surrounding molecular clouds, the maser emission arises 
in the post-shock gas \citep{lo05}.

The Megamaser Cosmology Project has the primary goal of measuring the 
Hubble Constant, H$_0$, by determining angular diameter distances to
circumnuclear megamasers in AGNs well into the Hubble flow \citep{reid09, bra10, reid13, kuo13, gao161}. 
A second important goal of the project is to measure precise masses of
the supermassive black holes (SMBHs) in the nuclei of these galaxies.
Very Long 
Baseline Interferometry (VLBI) provides both the sub-milliarcsecond
angular resolution needed to map the extremely compact masing regions
and the spectral resolution needed to study 
their kinematics. VLBI can resolve the gravitational 
``sphere of influence'' of the SMBH in many megamaser disks, and the 
mass can be measured accurately when the rotation of the edge-on
disk is observed to be Keplerian. {\bf To date,  with this method, 20 SMBH 
masses have been measured (e.g. \citealt{ggr96, hmg99, gbe03, kgm05, kgm08, reid09, kuo11, Yam2012, gao161, gre16a, gao17}.)} 
Most of the measured values are between $10^6$ and a few times 
of $10^7 M_\odot$, and the extremely high mass densities 
($\rho\gtrsim 10^{11} M_\sun \mathrm{~pc}^{-3}$) inside the maser 
disks argue strongly against alternatives to SMBHs (e.g., a dense 
cluster of stars or stellar remnants) as the central objects in 
these galaxies. The ``gold standard'' SMBH masses measured with megamasers provide
a strong test of the $M_\mathrm{SMBH}$--$\sigma_\star$ relation 
at its low-mass end \citep{fm00, gbr00, grg09, kor11, kbc11, gre12, gre16a, gre16b}.

Although edge-on disk masers in Keplerian rotation are the best-studied class of megamasers, not all megamaser sources originate from such systems.
Some megamasers form in non-Keplerian disks,
and some show complex geometries influenced by nuclear jets and outflows.
Still other megamasers have single-dish spectral profiles with hints of disk emission, 
such as high-velocity components, but their
VLBI maps reveal complex geometries and disordered dynamics. In these cases, 
SMBH masses can not be determined simply by fitting a Keplerian rotation curve. 

In this paper, we present VLBI images of megamasers whose single-dish
spectral profiles show a range of properties.  One has the characteristic
triple-peaked disk profile (UGC~6093). Others have
asymmetric high-velocity maser components, or broad systemic components
with no apparent high-velocity counterparts.
We determine an accurate SMBH mass for
UGC~6093 by fitting a disk model to the observed maser components.
For other systems with high-velocity components, we apply other methods to determine constraints on the
enclosed mass, assuming the masers originate in a disk.
Our mass estimates for the SMBHs in these megamaser systems are proportional to its distance,
which we calculate from the galaxy's recessional velocity assuming
$H_0 = 70 \mathrm{~km~s}^{-1}\mathrm{~Mpc}^{-1}$ \citep{dun09}.

\section{The Sample, Observations and Data Reduction}

\subsection{The Megamaser Sample}
The seven galaxies presented in this paper are Seyfert 2 galaxies 
and LINERs, whose H$_2$O megamasers were discovered between 
1994 and 2011. The H$_2$O masers in UGC~6093, J1346+5228 and 
J0350$-$0127 were discovered by the MCP survey with the Green Bank 
Telescope (GBT), while the remaining masers were discovered in 
earlier surveys \citep{Braatz94, Hagiwara97, Greenhill03}.
Table \ref{galaxy} lists the properties (e.g. name, recession 
velocity referenced to the local standard of rest (LSR), referred 
to as systemic velocity hereafter, angular size distance when 
$H_0 = 70\mathrm{~km~s}^{-1} \mathrm{~Mpc}^{-1}$, morphology 
and AGN type) of the galaxies and discovery references. 
Only NGC~5793 had prior VLBI images, but since only blueshifted 
masers had been detected \citep{Hagiwara01}, we re-observed 
NGC~5793 to image the redshifted and systemic masers
to test the ``compact molecular disk'' model proposed by Hagiwara.

\subsection{Observations}
{\bf We observed all seven galaxies between 2010 May and 2013 January 
with the Very Long Baseline Array (VLBA) augmented by the Green 
Bank Telescope (GBT), meanwhile the 100-m Effelsberg (EB) 
telescope was also used for UGC~6093.}
Mrk~1210 and UGC~6093 were observed for three tracks 
each; the other five galaxies were observed for one track each. 
Details of the observations are listed in 
Table \ref{Obs}. The observations were designed to minimize 
phase errors that could degrade the accuracy of the maser positions: 
(1) To remove the tropospheric delay and clock errors, we scheduled two
``geodetic blocks'' with durations of 45--60 minutes and observed in left
circular polarization with eight 16~MHz bands spanning 374 to 476~MHz 
(see the frequency coverage for each VLBI track in Table \ref{Obs}) 
before and after the ``maser blocks'' for target galaxies. 
In each geodetic block, $\sim$20 compact radio 
sources covering a wide range of zenith angles were observed to 
estimate vertical tropospheric delays and clock offsets at each antenna.
(2) To remove electronic delay differences among intermediate frequency (IF) sub-bands, 
scans of 5--10 minutes duration on strong compact sources 
were interspersed between the maser blocks. (3) All megamasers 
except J1346+5228 were observed in self-calibration mode, where 
a bright maser spot was used to remove phase variations from 
the atomic clocks and tropospheric delays (see LSR velocities 
of the masers used as calibrators in column 8 of Table \ref{Obs}).
J1346+5228 was observed in phase-referencing mode, where 
the telescope pointing was switched between the target maser 
and a nearby (within $1^{\circ}$) phase-reference source {\bf (4C~+53.28)} every 
50~s to provide phase calibration.

\subsection{Data Reduction}
We reduced the data using the NRAO Astronomical Image Processing
System (AIPS) following the standard procedures for the MCP project
\citep{reid09, kuo11, reid13, kuo13, gao161}. 
Then the calibrated interferometer $(u,v)$-data from adjacent 
VLBI tracks of Mrk~1210 (BK163A and BK163C, referred as BK163AC and 
considered as one track hereafter) and UGC~6093 (BB294C and BB294E, 
referred as BB294CE and considered as one track hereafter) 
were combined for better sensitivity. Next, we spectrally 
smoothed the calibrated data to 125~kHz ($\sim1.8\mathrm{~km~s}^{-1}$) 
per channel, then constructed images by Fourier transformed and CLEANing with the 
AIPS task IMAGR. Columns 5 and 6 of Table \ref{Obs} list the 
synthesized half-power beamwidths, position angles, and rms 
sensitivities of these images. We fit elliptical Gaussians to 
the detected maser features to obtain their positions and peak 
flux densities. Table \ref{masers} shows a sample of fitting 
results for VLBI track BB278I of UGC~6093, including nonrelativistic 
``optical convention'' radial velocities defined by $V \equiv c (\Delta \lambda /\lambda)$. 
We generate spectra from the VLBI cubes to compare with 
single-dish spectra. The VLBI spectra 
were produced by using the fitted peak flux densities and velocities of 
masers with SNR$>$3 (SNR is the ratio of the peak flux density 
to its rms uncertainty), while using the rms noise of the 
CLEANed VLBI image as the noise level of the spectra. 

To make the maser maps, we determined the position 
offsets of masers with SNR$>$5 relative to the brightest 
systemic maser, except in the case of NGC~5793 where we measured
offsets relative to the brightest redshifted maser.
%Further steps were taken for UGC~6093, Mrk~1210 and J~1346+5228. 
For UGC~6093, to refine the VLBI map and show the emission structure more 
clearly, we grouped the masers detected in its VLBI spectra into 
clumps, and averaged the data in the velocity range (see the velocity 
range of each clump detected in VLBI track BB278I in Table \ref{bb278i_bin}, 
and see the velocity range of each clump detected in 
BB294CE in on-line Table 4.1) of each clump spectrally to increase 
the SNR and thus decrease the position error. Then we Fourier transformed, CLEANed
and fitted the re-averaged data (see the fitting 
results of re-averaged data of VLBI track BB278I in Table \ref{bb278i_bin}, 
as well as fitting results of re-averaged data in BB294CE 
in on-line Table 4.1). We referred positions of masers detected 
in VLBI track BB278I and BB294CE to the masers in the velocity range 
$10834.67-10850.07\mathrm{~km~s}^{-1}$ and $10814.30-10850.52\mathrm{~km~s}^{-1}$ respectively,
and finally made a single map by aligning the positions of those two masers. 
Mrk~1210 also has two VLBI data sets (BB313B and BK163AC). 
We first referred positions of masers detected in both of them
to the maser at velocity $4227.84\mathrm{~km~s}^{-1}$, 
then made a single map by aligning the positions of the reference
maser, and finally moved the map center to the 
position of the only systemic maser. For J1346+5228, to refine the VLBI map,
we spectrally smoothed data containing systemic masers to 500~kHz ($\sim7.2\mathrm{~km~s}^{-1}$) per channel, 
and averaged channels containing red and blueshifted masers into 
two channels respectively (see the velocity range we used to 
average the data in on-line Table 4.1). We produced the final 
map of J1346+5228 with the re-averaged data (see the fitting 
results of re-averaged data of VLBI track BB313X in online Table 4.1).  

\section{VLBI Spectra and maps of H$_2$O Megamasers}

\begin{figure}
\centering
\includegraphics[angle=0, scale=0.3]{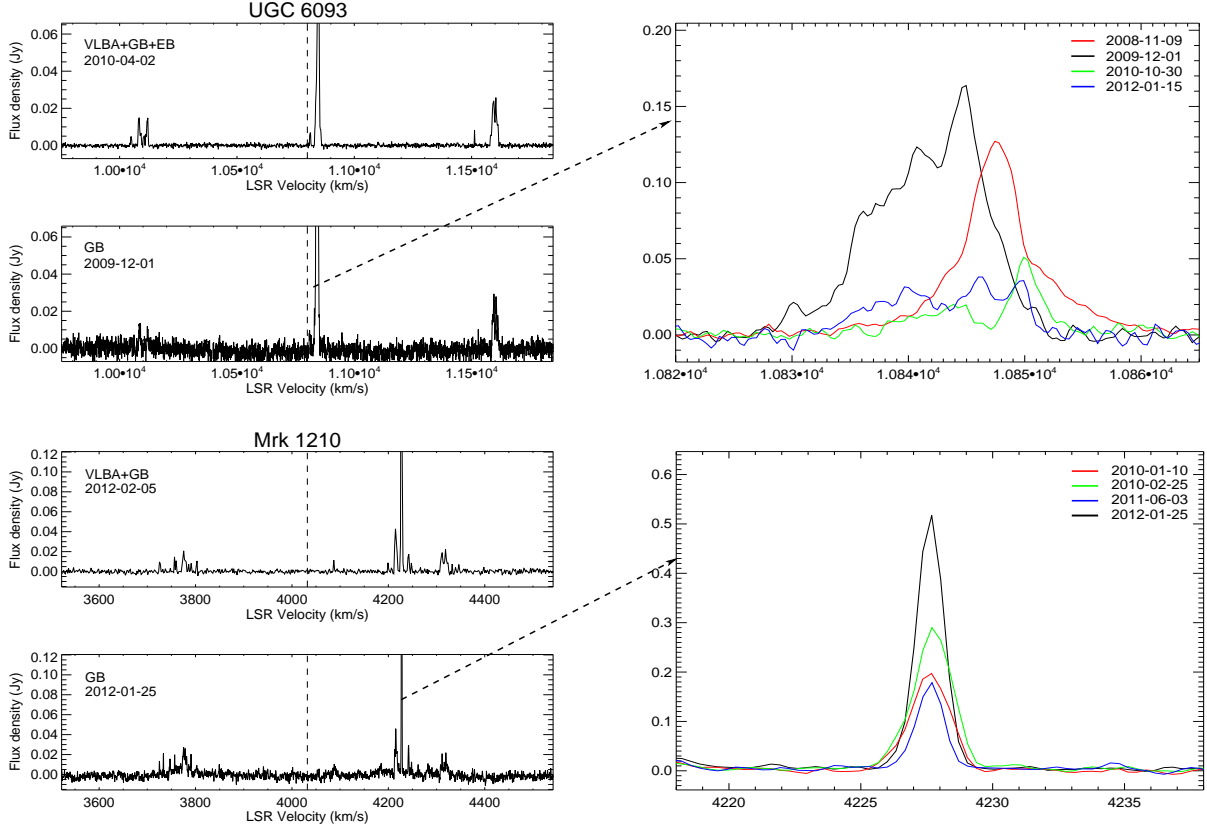}
\caption{The left panels show the VLBI spectra of megamasers in UGC~6093 and Mrk~1210, compared with quasi-simultaneous
high-sensitivity (rms$\sim 3\mathrm{~mJy~beam}^{-1}$) GBT spectra.
The vertical dashed lines denote the LSR ``optical'' velocities of the host galaxies from NED.
The right panels present the zoomed-in GBT spectra (black) overlapping
three other epochs (colors) during the monitoring period to show the variability of masers
in a certain velocity range.  For UGC~6093 
both the spectral profiles and flux densities of the systemic masers are 
quite variable; for Mrk~1210, a spectral line at $4228\mathrm{km~s}^{-1}$ kept 
flaring during its GBT monitoring period; e.g., between June 2011 and January 2012 the peak 
flux density of this maser grew from $190\mathrm{~mJy~beam}^{-1}$ to nearly $500\mathrm{~mJy~beam}^{-1}$.}
\label{spectraa}
\end{figure}

As mentioned in Section~2.3, we compared the VLBI spectra with 
quasi-simultaneous high-sensitivity ($\sigma\sim 3\mathrm{~mJy~beam}^{-1}$) 
GBT spectra in Figure \ref{spectraa}. For Mrk~1210 and UGC~6093, 
the most sensitive VLBI tracks (BB313B and BB278I for Mrk~1210 
and UGC~6093 respectively) were used in this comparison. 
Almost all maser features from the single-dish spectra are 
detected in our VLBI observations, except for one blueshifted 
maser and a small fraction of redshifted masers 
in NGC~6926, and one redshifted maser in NGC~5793. 
The VLBI flux densities of most masers agree well with the 
single-dish observations.  The differences between the VLBI 
and single-dish flux densities of some individual lines can mainly be
attributed to maser variability. 

UGC~6093 has been monitored occasionally by the MCP with the GBT since 2007. The upper left 
panel of Figure \ref{spectraa} compares the VLBI spectrum with 
a GBT spectrum taken four months earlier, on 2009 December 1. 
UGC~6093 has the characteristic triple-peaked spectrum of a disk maser. 
Compared to other disk maser systems, the red and blue-shifted masers
span relatively narrow velocity windows of $74\mathrm{~km~s}^{-1}$ and $92\mathrm{~km~s}^{-1}$, respectively. 
The upper-right panel of Figure \ref{spectraa} displays single-dish spectra of 
the systemic masers from four epochs, showing that the 
systemic masers are quite variable.

Mrk~1210 has been monitored with the GBT periodically since 2003. The bottom left panel 
of Figure \ref{spectraa} shows the GBT spectrum taken on 
2012 January 25 compared with the VLBI spectrum taken 11 days 
later. While Mrk~1210 does not have the characteristic triple-peaked 
spectrum of a disk maser, the spectrum does show a large overall range of detected velocities
covering $\sim$ $600\mathrm{~km~s}^{-1}$, bracketing the systemic velocity, and thus
suggestive of high-velocity rotation around a SMBH.
The redshifted maser detected near $4228\mathrm{~km~s}^{-1}$ with 
a width of $\sim3\mathrm{~km~s}^{-1}$ was flaring at the time 
of the observations. The weak near-systemic maser at $4088\mathrm{~km~s}^{-1}$ 
with a width of $1.7\mathrm{~km~s}^{-1}$ only appeared 
for a short period in early 2012. The bottom right panel of 
Figure \ref{spectraa} displays single-dish spectra on the 
flaring maser feature at four epochs. The peak flux density varied 
from $0.19\mathrm{~Jy~beam}^{-1}$ to nearly $0.5\mathrm{~Jy~beam}^{-1}$ 
within two years, while weaker high-velocity masers were relatively stable.

\begin{figure}
\centering
\includegraphics[angle=0, scale=0.6]{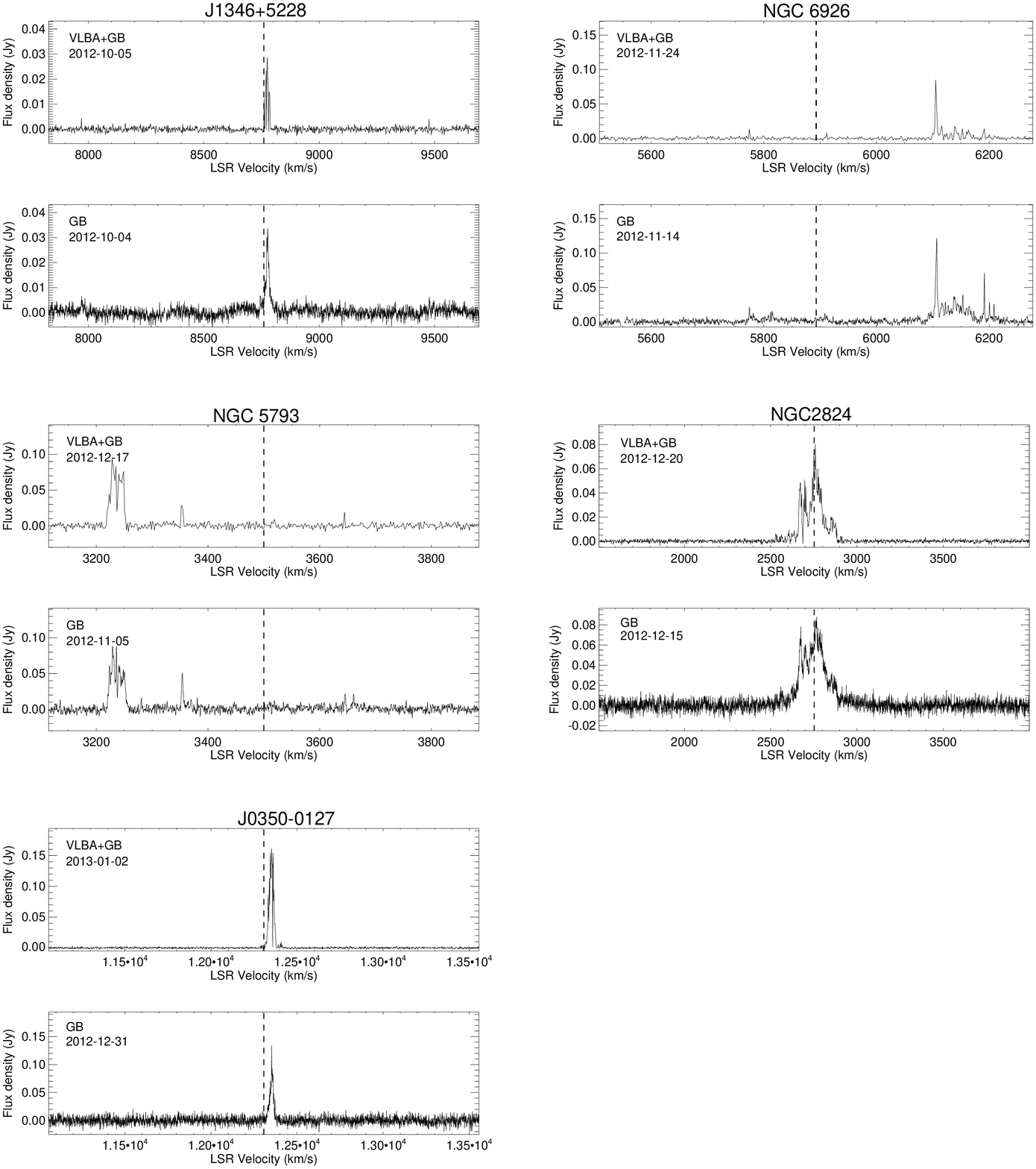}
\caption{VLBI spectra of megamasers in J1346+5228, 
NGC~6926, NGC~5793, NGC~2824, and J0350$-$0127 compared with quasi-simultaneous
high-sensitivity (rms$\sim 3\mathrm{~mJy~beam}^{-1}$) GBT spectra.
The vertical dashed lines denote the LSR ``optical'' velocities of the host galaxies from NED.}
\label{spectrab}
\end{figure}

J1346+5228 (Figure \ref{spectrab}) has the typical triple-peaked 
spectrum of disk masers, but the high-velocity masers are barely 
detectable at flux densities $<$10~mJy.  
The redshifted masers peaking around $9478\mathrm{~km~s}^{-1}$ 
and blueshifted masers peaking around $7972\mathrm{~km~s}^{-1}$
are symmetrically offset from the systemic velocity 
by more than $700\mathrm{~km~s}^{-1}$, {\bf and cover narrow velocity 
ranges of $6\mathrm{~km~s}^{-1}$ 
and $9\mathrm{~km~s}^{-1}$ respectively.}

NGC~6926 has a spectrum asymmetric in both velocity 
and strength (Figure \ref{spectrab}). 
The spectrum shows a very weak feature at 
$5912\mathrm{~km~s}^{-1}$  near the systemic velocity 
and a broad (width $>100\mathrm{~km~s}^{-1}$) redshifted line complex 
with flux densities over a few tens of mJy. Two blueshifted features with 
flux densities of 20~mJy and 15~mJy are detected at $5775\mathrm{~km~s}^{-1}$ 
and $5815\mathrm{~km~s}^{-1}$ in the single-dish spectrum
respectively, while only the former one is detected in the VLBI observation. 
 
Hagiwara discovered the H$_2$O megamaser in NGC~5793 with the Nobeyama 
45~m telescope \citep{Hagiwara97} and made subsequent interferometric studies \citep{Hagiwara01}.  
Single-dish observations made in the 1990s showed a hint of features near 
the systemic velocity. Blueshifted features were detected  $>200 \mathrm{~km~s}^{-1}$ 
away from the systemic velocity while redshifted features at $3677\mathrm{~km~s}^{-1}$ were detected in only 
one epoch by the Nobeyama 45~m. Our GBT and VLBI spectrum (Figure \ref{spectrab}) 
revealed no detected maser features near the systemic velocity but both the blueshifted and redshifted 
maser features were detected for the first time in the VLBI observation.

The single-dish and VLBI spectra of both NGC~2824 and J0350$-$0127
show no high velocity features characteristic of a disk maser.
Most of the megamaser emission from NGC~2824 comes from a broad line
complex in the velocity range from 2500 to $3000\mathrm{~km~s}^{-1}$ (Figure \ref{spectrab}). 
The broad profile of the spectrum of NGC~2824 is similar to 
those of NGC~1052 and Mrk~348, in which the 
masers originate from the pc-scale nuclear jet \citep{clau98, fal2000, pec03}. 
In J0350$-$0127, the masers are confined to a narrow velocity range between 
12320 and $12380\mathrm{~km~s}^{-1}$ and peak at $12352\mathrm{~km~s}^{-1}$, 
only $45\mathrm{~km~s}^{-1}$ from the systemic velocity.

\subsection{UGC~6093}

\begin{figure}
\includegraphics[angle=0, scale=0.68]{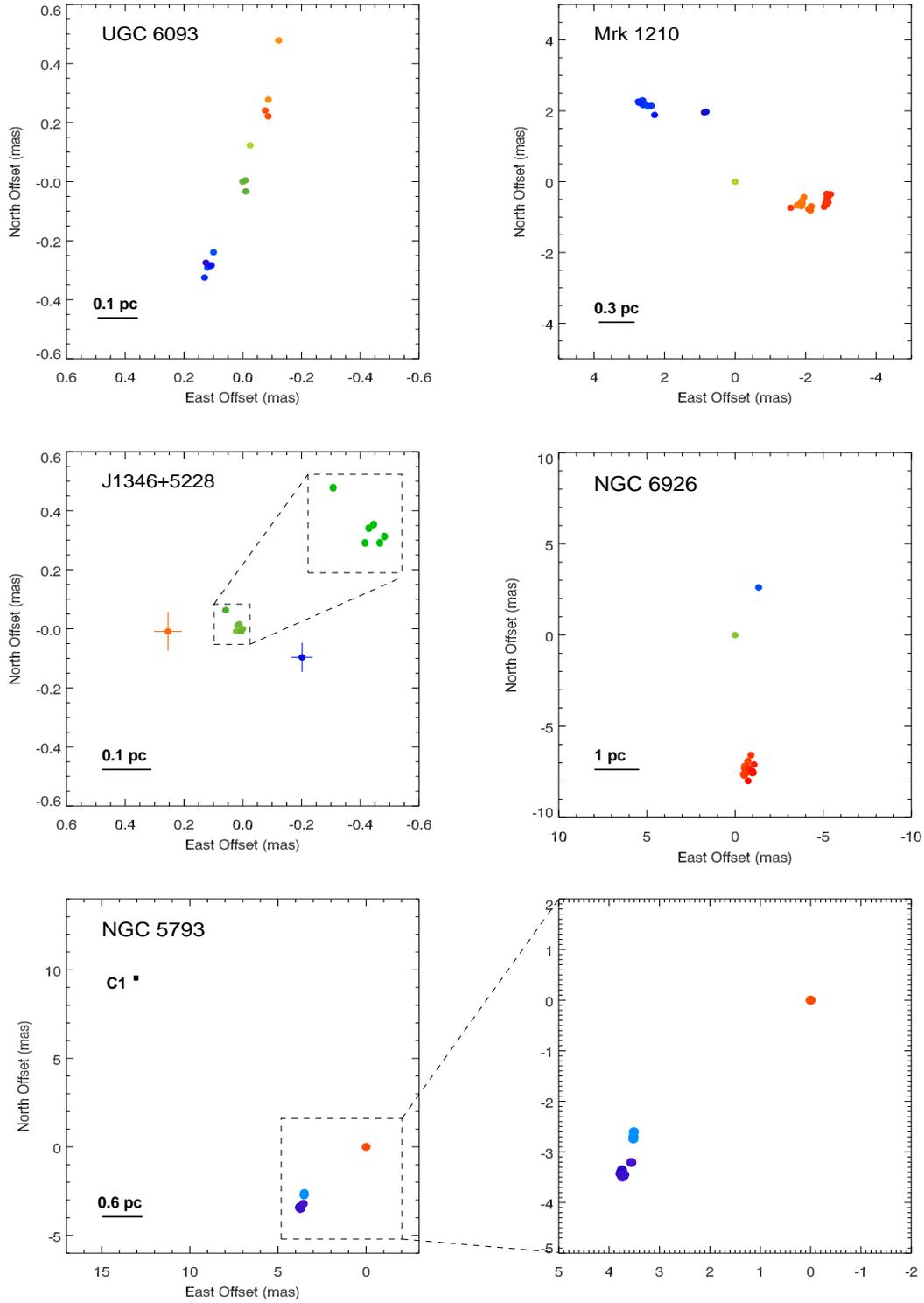}
\caption{\scriptsize{VLBI maps of UGC~6093, Mrk~1210, J1346+5228,
    NGC~6926, and NGC~5793. The data points in the maps are color-coded
    to indicate redshifted (red), blueshifted (blue), and systemic (green)
    masers. Except for NGC~5793, each map is centered
    on the position of the brightest systemic maser. NGC~5793
    has no detectable systemic masers, so its map is centered at the
    brightest redshifted maser feature. A continuum component with SNR $\sim$ 5 
    which is believed to be a part of the jet component
    C1 \citep{Hagiwara01} was detected $\sim 20 \mathrm{~mas}$ northeast
    of the maser emission of NGC~5793 (marked with a black square on the map of NGC~5793).}}\label{mapa}
\end{figure}

The VLBI map of UGC~6093 (Figure \ref{mapa}) reveals an edge-on disk. 
The well-aligned systemic masers and high-velocity masers define a disk 
plane with little or no warping. The redshifted side of the disk plane 
has a position angle $PA\sim -20^{\circ}$  east of north (this definition 
of position angle $PA$ is kept in the following subsections). The diameter 
of the disk is about 0.9 mas or $\approx 0.6\mathrm{~pc}$, which is within the typical range of sizes
for megamaser disks.
 
We used a Bayesian fitting program \citep{reid13} to estimate the mass 
of the central black hole in UGC~6093. The program was originally used 
to estimate the Hubble constant for MCP galaxies. By fitting the 
positions, radial velocities (with relativistic corrections), 
and accelerations (measured by monitoring velocity shifts of the 
systemic maser features) of masers, this program estimates the 
positions of masers in the disk plane and gives best-fit values 
for the mass and position of the SMBH, recession velocity of 
the SMBH, and the Hubble constant (an angular-size distance 
can be inferred from the latter two parameters). 

We monitored the maser spectrum of UGC~6093 periodically with the
GBT over nine years, but the systemic lines were too variable and 
blended to track velocity drifts reliably from epoch to epoch. 
So the maser in UGC~6093 is not well suited to a distance measurement 
for the MCP. Nevertheless, we can still apply the full 3D fitting 
program to measure the BH mass based on an adopted value of Hubble 
constant $H_0 = 70\mathrm{~km~s}^{-1}\mathrm{~Mpc}^{-1}$. We fit 
a flat disk because the VLBI map does not reveal a warp in a 
convincing way. We also fit circular orbits, since in previous 
MCP work \citep{reid13, kuo13, gao161}, any eccentricity for 
maser feature orbits turned out to be very small. So six parameters 
remained to be fit: three for the position and mass of the SMBH 
(the right ascension $x_0$, declination $y_0$, and mass $M_\mathrm{SMBH}$), 
two for the geometry of the disk (position angle $PA$ and inclination 
angle $i$), and one for the recession velocity referenced to 
the LSR ($v_\mathrm{LSR}$). We use only maser features with 
SNR $>$7. Since we did not have reliable acceleration, 
we adopted nominal acceleration values with large uncertainties 
($4\pm50\mathrm{~km~s^{-1}~yr^{-1}}$). The fit, then, is still 
constrained in part by the measured positions of those maser 
spots, since the mass measurement is almost free from the value of 
acceleration，even removing the systemic features from the fit
produced no significant changes to the black hole mass estimate.

\begin{figure}[ht]
\begin{center}
\includegraphics[angle=0, scale=0.7]{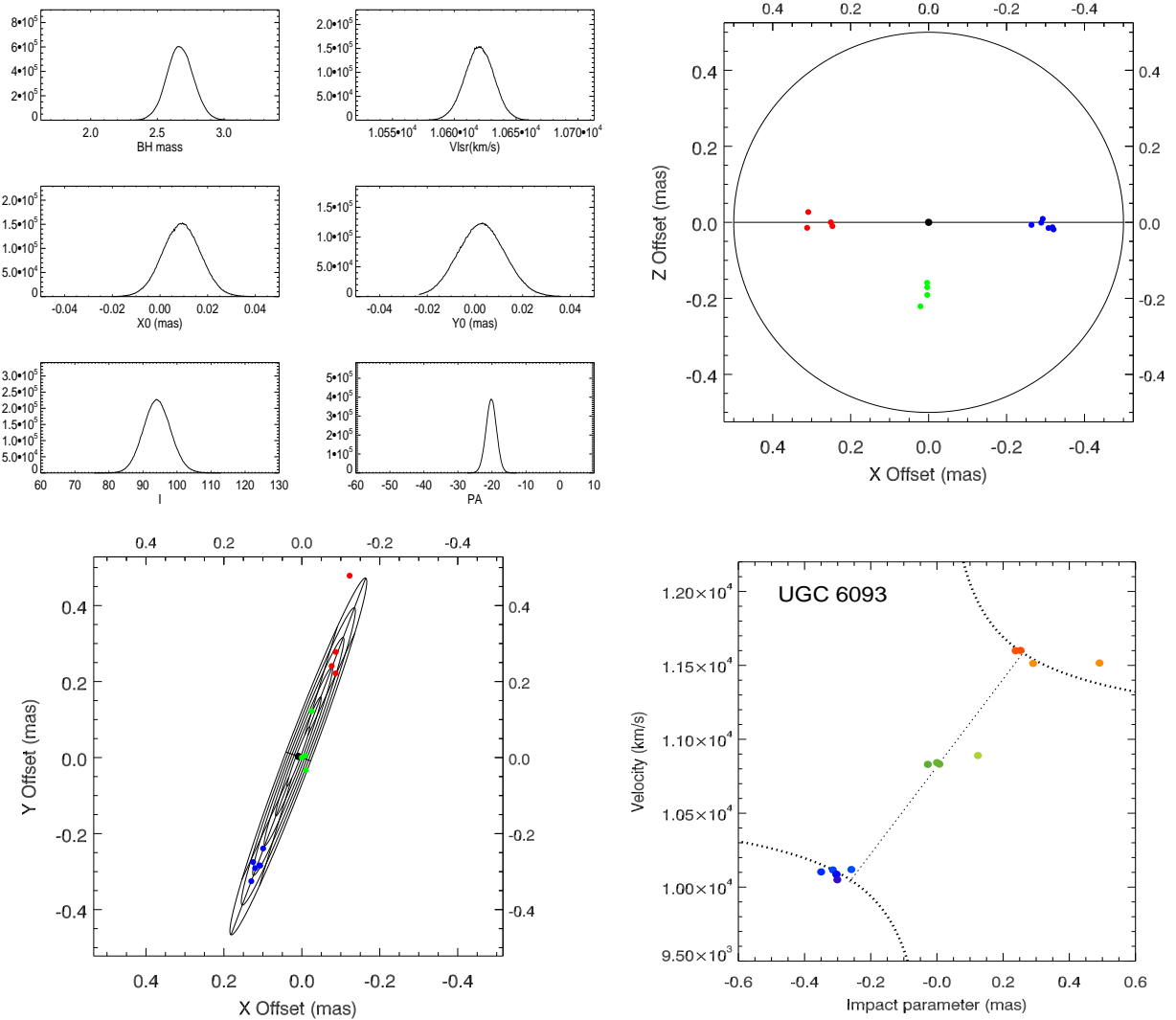}
\caption{Visualization of Bayesian fitting results for UGC~6093.
The upper left panel presents the probability density functions (PDFs)
of the fitted parameters, while the upper right panel shows 
a top view of the fitted maser positions. The lower left panel shows 
the 3-D disk model based on the fitting results overlapping with the 
observed positions of masers with SNR $>$ 7. The lower right panel
presents the $PV$ diagram, and the dash line 
present the position-velocity relation in an ideal case, 
an edge-on, flat, Keplarian disk orbiting around 
a point mass with our fitted value.}\label{UGC6093_baye}
\end{center}
\end{figure}

We list the most likely parameter values and the 68\% confidence 
ranges derived from Bayesian fitting for UGC~6093 in the first 
line of Table \ref{baye} and show the fitting results graphically 
in Figure \ref{UGC6093_baye}.  Marginalized probability density functions 
(PDFs) of the parameters (generated by binning the the MCMC trials) 
are presented in the upper left panel. These PDFs appear smooth and 
symmetric for all parameters. The upper right panel is a top view 
of the maser disk, showing the best-fit maser positions. The lower 
left panel shows the 3-D disk model superposed with masers as seen 
on the sky. The lower right panel presents the $PV$ diagram. The 
dashed line presents the position-velocity relation in an ideal case, 
an edge-on, flat, Keplerian disk orbiting around a point mass with 
our fitted value. As the figure shows, data points of UGC~6093 
slightly deviate from the ideal model.

The distance from the masers to the SMBH ranges between 0.17 and 0.33 mas (0.12 to
0.24 pc). Almost all of the high-velocity masers 
are located along the mid-line of the disk as is common for disk 
megamasers. The 3-D geometric model (an edge-on disk at inclination 
angle $i = 93.8^{\circ}$ and position angle $PA = -20.1^{\circ}$) 
constrained by our observations is a good fit. The SMBH lies 
within 0.01~mas of the reference center of the VLBI map, and the 
fitted mass is $M_\mathrm{SMBH} = 2.58^{+0.11}_{-0.09}\times10^7 M_\odot$.
For the black hole mass, in addition to the formal fitting error 
($\sim 3.8\%$) there are two main sources of systemic error: 
{\bf (1) Uncertainty in the distance to the galaxy. In this paper, 
we adopted $H_0 = 70 \mathrm{~km~s}^{-1} \mathrm{~Mpc}^{-1}$ with 
an uncertainty of 5\% ($\sigma_1$). This uncertainty will pass down 
to the mass, which is proportional to $H_0$. (2) Uncertainty in 
the absolute positions for the reference maser spots, which
are $\sim 10 \mathrm{~mas}$ based on Very Large Array measurements;
for self-calibrated VLBI observations, this position uncertainty
introduces an error up to 5\% ($\sigma_2$) in the SMBH mass \citep{kuo11}. 
Combining these sources of systematic uncertainty, we estimate
a total systematic uncertainty in the SMBH mass of $\sqrt{\sigma_1^2+\sigma_2^2}\sim7\%$.}
The best fit LSR recession velocity of the SMBH is 
$v_\mathrm{LSR}=10621^{+11}_{-12}\mathrm{km~s}^{-1}$. 
An angular-size distance $d_A=151.7\mathrm{~Mpc}$ can be 
inferred from the velocity of the SMBH and our adopted 
Hubble constant. But we do not know the exact value of 
angular-size distance to UGC 6093, so specifying $M_\mathrm{SMBH}$ in terms of $D_A$ 
is a good compromise. Since our $M_\mathrm{SMBH}$ estimates 
are directly proportional to the angular-size distance, 
the mass can be expressed as $M_\mathrm{SMBH}=2.58\times10^7 \Biggl(\frac{D_A}{151.7\mathrm{~Mpc}}\Biggr)M_\odot $.

\subsection{Mrk~1210}

\begin{figure}[ht]
\begin{center}
\includegraphics[angle=0, scale=0.7]{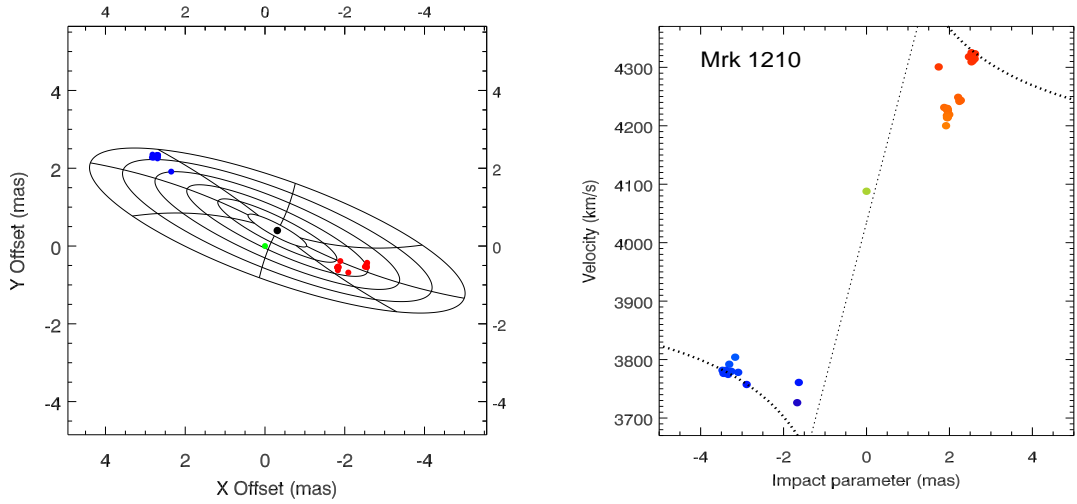}
\caption{The left panel of shows a concept map of a warped and tilted disk overlapping with the observed masers with SNR$>$7 in Mrk~1210. The right panel shows the $PV$ diagram, in which the dash line presents the $PV$ relation of an edge-on, flat, Keplarian disk orbiting around a point mass of $1.42\times10^7M_\odot$.}\label{mrk1210_baye}
\end{center}
\end{figure}

The Mrk~1210 VLBI map in Figure \ref{mapa} shows 
redshifted and blueshifted maser features in separate clusters 
separated by 7~mas or $\approx 2 \mathrm{~pc}$, with a $PA \sim -120^{\circ}$.  
A single maser component detected near the systemic velocity falls
between the two clusters but offset from the connecting line.
While the general configuration of maser features is roughly consistent
with disk rotation, the exact nature of the maser system remains inconclusive.
If indeed that masers arise in a disk, the
misalignment of the maser features suggests the potential of warping or
eccentricity in the disk, or an inclination different from edge-on.  The left panel of 
Figure \ref{mrk1210_baye} shows a concept map of a warped and tilted disk overlapping 
with the observed masers with SNR$>$7 in Mrk~1210. 

Assuming a disk origin and approximating the radius as half the full extent of the megamaser emission,
and the rotation velocity as the velocity offset of the high-velocity masers from the systemic velocity,
we estimated the mass enclosed within each high-velocity maser component. These values are distributed 
over a wide range from 0.2 to 2.2$\times10^7 M_\odot$.
We also tried to estimate the SMBH mass using the same 3-D model fitting method that we applied to UGC~6093. 
Some parameters (e.g. $x_0$ and $y_0$) did not converge in the attempted fit, implying the present data do not adequately constrain the 
assumed model.  We consider a representative value for the SMBH mass ($1.42\times10^7 M_\odot$), 
which we use for reference only in Figure\ref{mrk1210_baye} and Table \ref{baye}.
The dashed line in the right panel of Figure\ref{mrk1210_baye} presents the $PV$ relation of an edge-on, flat, Keplarian disk 
orbiting a point mass with the reference value. Most masers, especially the red-shifted ones with velocities below $4400\mathrm{~km~s}^{-1}$ and 
including the flaring feature detected around $4228\mathrm{~km~s}^{-1}$, obviously deviate from the fit.  These deviations represent
unmodeled complexity in this maser system.

\subsection{J1346+5228}

\begin{figure}[ht]
\begin{center}
\includegraphics[angle=0, scale=0.55]{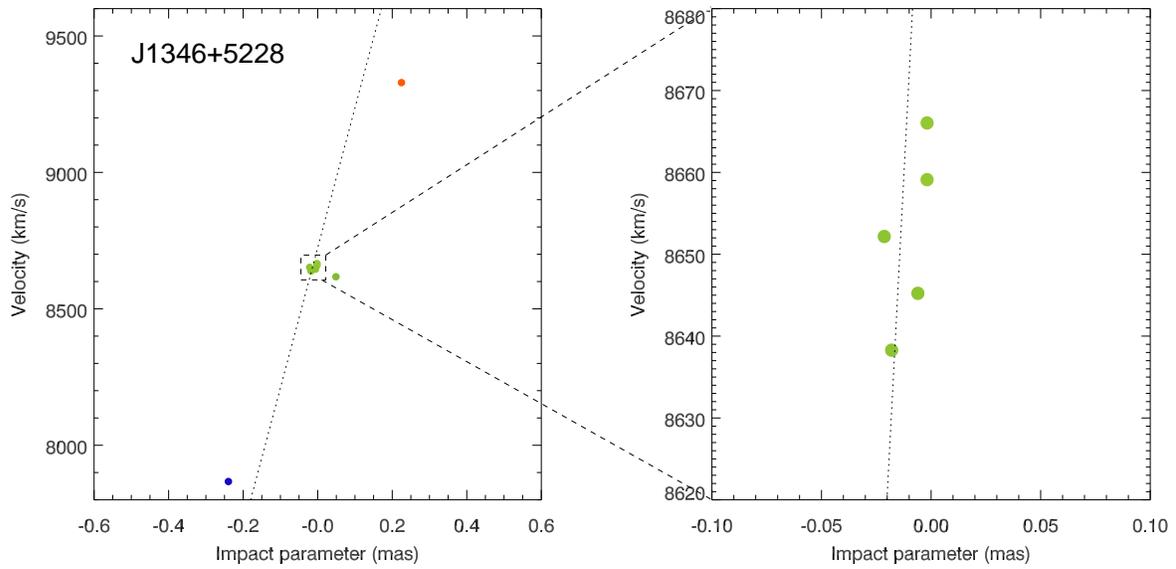}
\caption{Position–-Velocity ($PV$) diagram of J1346+5228.
The offsets between the  positions of masers projected onto the 
fitted disk plane ($PA = 77.3^{\circ}$) and the unweighted 
average position of systemic masers are defined as the "impact parameters".
The distribution of five (out of a total of six) systemic features
on the $\textsl{PV}$ diagram is consistent with masers at a
same radial distance from the dynamical center.}\label{J1346AVER}
\end{center}
\end{figure}

In the VLBI map for J1346+5228 (Figure \ref{mapa}), the high-velocity 
masers are only marginally detected (5$\sigma$ for the redshifted and 
4$\sigma$ for the blueshifted maser). The distribution of masers is 
consistent with a disk of diameter 0.5~mas or $\approx 0.3 \mathrm{~pc}$. 
Applying the ``edge-on'' disk assumption, we defined
the line going through the red and blue points as the disk plane, with
position angle $PA \approx 79.2^{\circ}$. We also approximated the location of the dynamic 
center as the unweighted average position of the six systemic masers.
The offsets between the projected positions of masers and the dynamic center 
on the disk plane are defined as the ``impact parameters''.  
These impact parameters and maser velocities with both special 
and general relativistic corrections were adopted to 
produce the $PV$ diagram shown in Figure \ref{J1346AVER}.

We made two mass estimates for J1346$+$5228:
(1) We used the impact parameters of the redshifted maser ($0.24
\mathrm{~mas} \approx 0.14 \mathrm{~pc}$) and the blueshifted maser
($0.22 \mathrm{~mas} \approx 0.13 \mathrm{~pc}$) and their velocity
offsets from the recession velocity $V_\mathrm{LSR}$ to estimate the enclosed mass. The
results are $1.5 \times 10^7 M_\odot$ and $2.0 \times 10^7 M_\odot$, 
respectively. (2) We noticed that five of the six systemic masers
line up well on the $PV$ diagram (see the enlarged inset in Figure \ref{J1346AVER}),
indicating that they form at a common radius from the dynamical center.  
For a flat edge-on Keplerian disk, the ratio of $P$ to $V$ is the orbital 
angular velocity $\omega$. A linear fit to the $PV$ diagram
of the five systemic masers yields $\omega \approx 5191
\mathrm{~km~s}^{-1} \mathrm{~mas}^{-1}$. If we assume that the
radial distance $r$ of the five systemic masers from the SMBH is about the same
as the radius to the high-velocity masers, the enclosed
mass $M=\omega^2r^3G^{-1}$ is between 1.5 and $1.9\times 10^7M_\odot$. 
This is within the range of masses enclosed by the high-velocity masers, 
adding confidence to our estimate that the SMBH in J1346+5228 is between
$ 1.5 \times10^7M_\odot$ and $2.0 \times10^7M_\odot$.

\subsection{NGC~6926}
The VLBI map for NGC~6926 in Figure \ref{mapa} reveals a complex, bent geometry.
The redshifted maser complex appears on the map as a cluster with no organized velocity structure.
We do not detect any masers with radial velocities $>6192 \mathrm{~km~s}^{-1}$. 
Only one blueshifted maser at 
$5775 \mathrm{~km~s}^{-1}$ was detected in the VLBI map, 
while the other at $5815\mathrm{~km~s}^{-1}$ seen in the GBT spectrum 
10 days earlier than the VLBI observation was not detected.
The position of the weak systemic maser is closer to the blueshifted 
maser than to the redshifted masers. This is unexpected since,
as the VLBI and single-dish spectrum shows, redshifted masers have 
greater rotation speed than the blueshifted masers. Whether the
"systemic" maser is a part of a potential disk maser system is not clear. The extent of the maser spots is about 11~mas or 4.4~pc, which is larger
than the typical diameters of other megamaser disks.  Although the observations do not clearly identify a disk origin in this case,
we still make an estimation of the NGC~6926 SMBH mass based on a flat, edge-on Keplerian-disk assumption. We estimate a disk radius
$r=2.2\mathrm{~pc}$ as half the extent of the megamaser emission, 
and use the largest velocity offset of the high-velocity masers 
from the systemic velocity ($309\mathrm{~km~s}^{-1}$ for NGC~6926) 
as the rotation velocity ($V_\mathrm{rot}$) to estimate an upper 
limit to the enclosed mass:

\begin{equation}
M_\mathrm{SMBH} \lesssim
4.8 \times 10^7 \Biggl(\frac{r}{2.2 \mathrm{~pc}}\Biggr)
\Biggl(\frac{v_\mathrm{rot}}{309 \mathrm{~km~s}^{-1}} \Biggr)^2\Biggl( \frac{D_A}{82.2\mathrm{~Mpc}} \Biggr) M_\odot
\end{equation}

\subsection{NGC~5793}

The multi-epoch VLBI observations of \citet{Hagiwara01} between 1996 
and 1998 detected only the blueshifted masers  distributed in two 
clusters separated by 0.7~mas, and no discernible velocity structure 
was found in these clusters. The authors proposed a circumnuclear ``compact molecular disk'' 
as a possible origin of the masers based on spectral line observations 
of H$_2$O and other molecules in the nuclear region of NGC~5793, and 
continuum observations of its pc-scale jets, even without the detection 
of any redshifted masers.

Our VLBI map in Figure \ref{mapa} reveals two blueshifted maser 
clusters, also without any discernible velocity structure. The cluster 
centered around $3355\mathrm{~km~s}^{-1}$ (marked with light blue dots) 
is 0.8~mas away from the other (dark blue). We also detected a redshifted 
maser at radial velocity $3645\mathrm{~km~s}^{-1}$ at $\approx5.5$~mas 
northwest of the blueshifted masers. A tentatively-detected 22~GHz continuum source 
with SNR $\sim$ 5 was detected $\sim 20\mathrm{~mas}$ northeast of 
the maser emission, and it may be a part of the 
pc-scale jet component C1 mentioned in Hagiwara's study.

Our redshifted maser was 
detected at the receding side of the disk, as Hagiwara's model predicted; 
our detection thus supports the "disk" scenario as a potential origin of the masers. 
The positions of the high-velocity masers define a disk plane at position angle 
$PA\sim -43^\circ$. For a flat, edge-on Keplerian-disk with radius 
of $r\approx3\mathrm{~mas}\approx0.7$~pc and rotation velocity $V \approx 279 \mathrm{~km~s}^{-1}$, 
an upper limit for the enclosed mass is:

\begin{equation}
M_\mathrm{SMBH}\lesssim
1.3 \times 10^{7} \Biggl( \frac{r}{0.7\mathrm{~pc}} \Biggr)
\Biggl(\frac{v_\mathrm{rot}}{279 \mathrm{~km~s}^{-1}} \Biggr)^2 \Biggl( \frac{D_A}{49.3\mathrm{~Mpc}} \Biggr)M_\odot
\end{equation}

This value is typical for SMBHs in maser galaxies, and it is one order
of magnitude lower than the result given in \citet{Hagiwara97}, in
which the scale of the disk had been constrained only by the distribution
of the 21-cm continuum emission.

\subsection{NGC~2824 and J0350$-$0127}

\begin{figure}
\includegraphics[angle=0, scale=0.7]{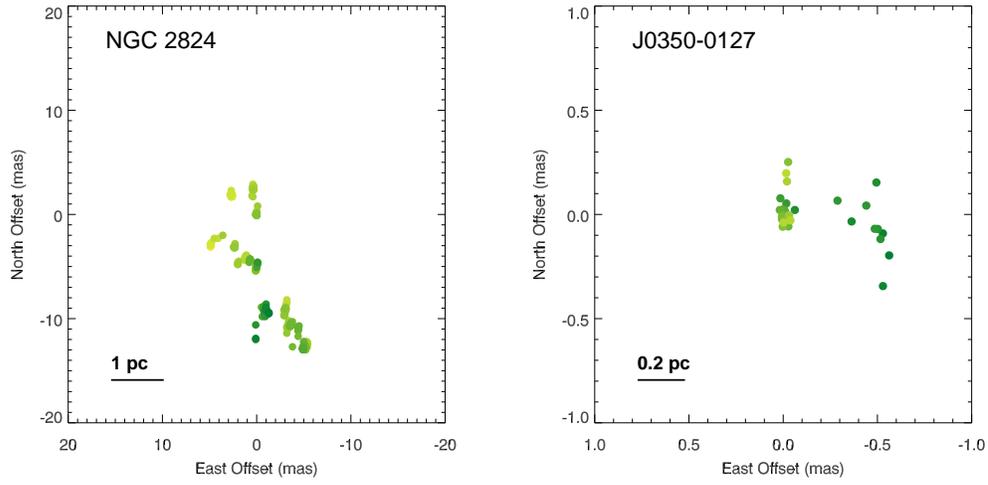}
\caption{VLBI maps of NGC~2824 and J0350$-$0127.  Each map is
  centered on the position of the brightest maser feature.}\label{mapb}
\end{figure}

On the VLBI map of NGC~2824 (Figure \ref{mapb}), the masers are distributed
over an area spanning 20~mas $\approx 3.8\mathrm{~pc}$. 
Although we do not detect continuum emission in this galaxy, the 
"broad line" spectral profile is reminiscent of pc-scale
jet masers in NGC~1052 and Mrk~348 \citep{clau98, fal2000, pec03}.

For J0350$-$0127, most of the  masers are clustered near the map center, 
while a few extend to the north by about 0.2~mas $\approx 0.2\mathrm{~pc}$, 
and others to the west by about 0.5~mas ($0.5 \mathrm{~pc}$). 
The origin of these masers is unclear.  They may reveal part of a maser disk, or 
they may be associated with diffuse outflows.

\section{Summary}

We present VLBI maps of H$_2$O megamasers for seven active 
galaxies, six for the first time. 

The H$_2$O megamaser in UGC~6093 clearly originates in a circumnuclear disk 
orbiting a SMBH of mass $M_\mathrm{SMBH} = 2.58\times10^7 M_\odot$ ($\pm 7\%$).
The H$_2$O megamasers in J1346+5228 are also in a disk, 
but in this case the high-velocity masers are extremely faint. We estimate its SMBH mass 
lies between 1.5 and $2.0 \times 10^7 M_\odot$. 

The maser systems in the other galaxies are complex and have uncertain origins.
In Mrk~1210, the spectral profile shows high-velocity components roughly symmetric
about the recession velocity of the galaxy.  We present a tilted and warped disk 
as one possible model to describe the maser complex, which extends over about 2 pc.
If the masers indeed originate in a disk, the SMBH mass is likely between 
0.2 and $2.2 \times 10^7 M_\odot$. 
The VLBI maps of NGC~6926 and NGC~5793 also reveal emission regions possibly 
associated with circumnuclear disks. Assuming disk origins, we estimated upper limits to the masses 
enclosed within the maser disks of $<4.8 \times 10^7 M_\odot$ for NGC~6926
$<1.3 \times 10^7 M_\odot$ for NGC~5793, Megamasers in the remaining two galaxies, 
NGC~2824 and J0350$-$0127, may be associated with $\sim$ pc-scale jets or outflows.

\acknowledgments This research has made use of the NASA/IPAC
Extragalactic Database (NED), which is operated by the Jet Propulsion
Laboratory, California Institute of Technology, under contract with
the National Aeronautics and Space Administration.  We thank the
referee for helpful comments, and we are grateful to the NRAO VLBA and
GBT staff for their many contributions to the Megamaser Cosmology
Project. The National Radio Astronomy Observatory is a facility of the
National Science Foundation operated under cooperative agreement by
Associated Universities, Inc. This work was partially carried out within the
Collaborative Research Council 956, subproject
A6, funded by the Deutsche Forschungsgemeinschaft (DFG)

{\it Facilities:} \facility{Effelsberg}, \facility{GBT}, \facility{VLBA}

\begin{table}
\caption{The Megamaser Sample}
 \label{galaxy}
\begin{tabular}{lrrlll}
\tableline\tableline\\
Name              & $V_\mathrm{LSR}^a$\qquad & $D_\mathrm{A}^b$ & ~Morphology $^c$ & AGN Type \rlap{$^d$} & Discovery Reference  \\
                  &(km~s$^{-1}$\rlap{)}      & Mpc &                  &                      & \\\tableline
UGC~6093          &10800                     & 147.8& ~SAB(rs)bc       & --                  & MCP, 2007 Dec 30 $^e$  \\
Mrk~1210          &4032                      & 56.7 & ~S?              & Sy                   & \cite{Braatz94}     \\
J1346+5228$^f$    &8760                      & 120.8& ~S0              & Sy2                  & MCP, 2006 Dec 12 $^e$ \\
NGC~6926          &5893                      & 82.2 & ~SB(s)bc pec     & Sy2                  & \cite{Greenhill03} \\
NGC~5793          &3500                      & 49.3& ~Sb? edge-on     & Sy2                  & \cite{Hagiwara97}  \\
NGC~2824          &2752                      & 38.9 & ~S0              & Sy                   & \cite{Greenhill03} \\
J0350$-$0127$^g$  &12307                     & 167.4& ~--              & --                   & MCP, 2011 Jan 4 $^e$ \\
\tableline
\end{tabular}
\noindent

\scriptsize{Notes:\\
$^a$ LSR ``optical'' velocity of host galaxy from the NASA/IPAC Extragalactic Database (NED). \\
$^b$ The angular size distance when $H_0 = 70 \mathrm{~km~s}^{-1}\mathrm{~Mpc}^{-1}$.\\
$^c$ Host galaxy morphological type (NED).  \\
$^d$ AGN type (NED) \\
$^e$ Megamaser discovery reference and date.\\
$^f$ NED name SBS 1344+527 \\ 
$^g$ NED name 2MFGC 03185 \\} 
\end{table}
\clearpage
 
\begin{sidewaystable}
\begin{center}
\caption{Observing Parameters}
 \label{Obs}
 \resizebox{210mm}{29mm}{
\begin{tabular}{llllrcccccc}
\tableline\tableline
 Experiment   & Date         &Galaxy        & Antennas~$^a$  & 
Synthesized Beam\rlap{~$^b$}     &Sensitivity~$^c$    & Observing~$^d$ & $V_{ref}^e$& Frequency coverage of  \\
    code      &              &                   &               & (mas $\times$ mas, deg)  & (mJy~beam$^{-1}$) & Mode & (km~s$^{-1}$\rlap{)} &   geodetic blocks (MHz)\\\tableline
BB278I        & 2010 Apr 2   &UGC~6093     & VLBA, GB, EB  & 1.00$\times$0.29, $-$11.63 & 0.84              & Self-cal. & 10847 & 21373.49-21781.49\\
BB294C        & 2011 Dec 17  &UGC~6093     & VLBA, GB, EB  & 1.20$\times$0.25, $-$15.13\rlap{$^f$} & 1.22\rlap{$^g$}      & Self-cal. & 10838 &21373.49-21781.49\\
BB294E        & 2012 Jan 29  &UGC~6093     & VLBA, GB, EB  &                          &                   & Self-cal. & 10838& 21373.49-21781.49 \\
BK163A        & 2010 May 22  &Mrk~1210     & VLBA          & 1.16$\times$0.42, $-$17.69\rlap{$^h$} & 1.25\rlap{$^i$}      & Self-cal. & 4230& 21800.49-22276.49\\
BK163C        & 2010 Jul 6   &Mrk~1210     & VLBA          &                          &                   & Self-cal. & 4230 & 21800.49-22276.49\\
BB313B        & 2012 Feb 5   &Mrk~1210     & VLBA, GB$^j$  & 1.26$\times$0.33, $-$12.24 & 1.94              & Self-cal. & 4229&21810.00-22184.00\\
BB313X        & 2012 Oct 05  &J1346+5228  & VLBA, GB      & 0.64$\times$0.45, $+$13.95  & 2.07              & Phase-ref \rlap{$^k$}. & --&21373.49-21781.49\\
BB313AE       & 2012 Nov 24  &NGC~6926     & VLBA, GB      & 1.26$\times$0.42, ~~$-$9.42  & 2.22              & Self-cal. & 6107 &21600.49-21981.49\\
BB313AI       & 2012 Dec 17  &NGC~5793     & VLBA, GB      & 0.98$\times$0.31, $-$12.02 & 3.18              & Self-cal.&3230& 21810.00-22286.00\\
BB313AJ       & 2012 Dec 20  &NGC~2824     & VLBA, GB      & 1.53$\times$0.42, ~~$-$8.29  & 0.69              & Self-cal. & 2761&21808.00-22284.00\\
BB313AL       & 2013 Jan 02  &J0350$-$0127 & VLBA, GB      & 1.12$\times$0.43, ~~$-$4.17  & 1.19              & Self-cal. &12351&21308.00-21784.00\\
\tableline
\end{tabular}}
\end{center}
\scriptsize{Notes:\\
$^a$ VLBA: Very Long Baseline Array; GB: Green Bank Telescope of NRAO; \\
     EB: Max-Plank-Institut f\"ur Radioastronomie 100m antenna in Effelsberg, 
     Germany.\\
$^b$ Except for BB313X, this column shows the restoring beam FWHM 
    size and position 
     angle produced with data averaged over frequency 
     channels used for self-calibration.  For BB313X, this column shows the 
     restoring beam FWHM size and position angle (degrees east or north)
     produced with data averaged over frequency channels that contain the brightest systemic 
     line feature.\\
$^c$ Except for the galaxy J1346+5228, this column shows the rms noise level 
     of the clean image produced with data averaged over frequency channels 
     used for self-calibration.  For BB313X, this column shows the rms noise 
     of the clean image produced with data averaged over frequency 
     channels that contain the brightest systemic line feature.\\
$^d$ ``Self-cal.'' means the ``self calibration'' mode is used.\\
     ``Phas-ref.'' means the ``phase-referencing'' mode is used.\\
$^e$ LSR velocities of the masers used as calibrators\\
$^f$ The average restoring beam FWHM  size and position angle
     produced from the data set combined with BB294C and BB294E\\
$^g$ The sensitivity of the cleaned image produced from the data set 
     combined with BB294C and BB294E}
$^h$ The average restoring beam  FWHM size and position angle of the image
     produced from the data set combined with BK163A and BK163C\\
$^i$ The sensitivity of the cleaned image produced from the data set 
     combined with BK163A and BK163C\\
$^j$ GB was in the array during the observation but is flagged in data 
     reduction due to its absolute phase instability.\\
{\bf $^k$ The phase-reference~source~is~4C~+53.28.}\\
\end{sidewaystable}

\clearpage
\begin{table}
\begin{center}
\caption{Sample maser fits for UGC~6093 from project BB278I}
\label{masers}
\begin{tabular}{ccccr}\\\tableline\tableline
$V~^a$         & $\Delta~V~^a$   & $\theta_x~^b$ & $\theta_y~^b$ & 
$F_\nu^c$\qquad\qquad \\
(km~s$^{-1}$)      &(km~s$^{-1}$)  & (mas) & (mas) & (mJy~beam$^{-1}$) \\\tableline
\tableline
10859.57 &1.82     &\llap{$-$}0.0063  $\pm$ 0.0261 & 0.0110 $\pm$ 0.0736 & 
3.98  $\pm$ 0.71\\
10857.76 &1.82     &\llap{$-$}0.0098  $\pm$ 0.0133  &  0.0224  $\pm$ 0.0375 &
8.27  $\pm$ 0.75\\
10855.95 &1.82     & 0.0485  $\pm$ 0.0127   &\llap{$-$}0.0789  $\pm$ 0.0358 &
8.77  $\pm$ 0.76\\
10854.14 &1.82     &\llap{$-$}0.0094  $\pm$ 0.0108   & 0.0215  $\pm$ 0.0304 &
10.20  $\pm$ 0.75\\
10852.33 &1.82     & 0.0026 $\pm$ 0.0033   &\llap{$-$}0.0057  $\pm$ 0.0094 &
35.41  $\pm$ 0.80\\
10850.52 &1.82     &\llap{$-$}0.0029  $\pm$ 0.0017   &\llap{$-$}0.0047  $\pm$ 0.0048 &
79.16  $\pm$ 0.92\\
10848.71 &1.82     &\llap{$-$}0.0019  $\pm$ 0.0013   &\llap{$-$}0.0003  $\pm$ 0.0037 &116.17  $\pm$ 1.05\\
10846.89 &1.82     &\llap{$-$}0.0000  $\pm$ 0.0013   & 0.0000  $\pm$ 0.0037 &
131.71  $\pm$  1.17\\
10845.08 &1.82     &\llap{$-$}0.0038  $\pm$ 0.0015   &\llap{$-$}0.0001  $\pm$ 0.0042 & 102.13  $\pm$ 1.03\\
10843.27 &1.82     & 0.0012 $\pm$ 0.0020   &\llap{$-$}0.0031  $\pm$ 0.0055 &
72.51  $\pm$ 0.97\\
10841.46 &1.82     &\llap{$-$}0.0038  $\pm$ 0.0024   &\llap{$-$}0.0095  $\pm$ 0.0067 & 53.06  $\pm$ 0.85\\
10839.65 &1.82     &\llap{$-$}0.0016  $\pm$ 0.0033   &\llap{$-$}0.0001  $\pm$ 0.0094 &
35.42  $\pm$ 0.80\\
10837.84 &1.82     &\llap{$-$}0.0013  $\pm$ 0.0040   & 0.0042  $\pm$ 0.0113 &
28.64  $\pm$ 0.78\\
10836.03 &1.82     &\llap{$-$}0.0014 $\pm$ 0.0050   & 0.0092  $\pm$ 0.0140 &
23.47  $\pm$ 0.79\\
10834.22 &1.82     & 0.0173  $\pm$ 0.0095   & 0.0226  $\pm$ 0.0269 &
11.72  $\pm$ 0.76\\
10832.41 &1.82     &\llap{$-$}0.0191  $\pm$ 0.0139   &\llap{$-$}0.0130  $\pm$ 0.0391 &
7.97  $\pm$ 0.75\\
10814.30 &1.82     & 0.0102  $\pm$ 0.0151   &\llap{$-$}0.0558  $\pm$ 0.0427 &
6.88  $\pm$ 0.71\\
10812.49 &1.82     &\llap{$-$}0.0019  $\pm$ 0.0161   &\llap{$-$}0.0497  $\pm$ 0.0455 &
6.61  $\pm$ 0.73\\
10810.68 &1.82     &\llap{$-$}0.0520  $\pm$ 0.0208   &\llap{$-$}0.0459  $\pm$ 0.0587 &
5.31  $\pm$ 0.75\\\tableline
\end{tabular}
\end{center}
\scriptsize{Notes:\\
$^a$ Velocities and channel widths in 
     km~s$^{-1}$ 
referenced to the local standard of rest (LSR) and 
using the nonrelativistic optical velocity convention.\\
$^b$ East–-west and north–-south position offsets from 
     the position of the brightest systemic maser.  The listed rms position 
     uncertainties reflect fitting random errors only. \\
$^c$ Fitted peak intensity and its rms uncertainty.
}
\end{table}
\clearpage

\begin{table}
\begin{center}
\caption{Re-averaged sample data for UGC~6093}
\label{bb278i_bin}
\begin{tabular}{ccccr}
\tableline\tableline\\
Velocity Range$^a$ &Velocity center~$^b$ & $\theta_x~^c$ & $\theta_y~^c$ & 
SNR~\rlap{$^d$} \\
(km~s$^{-1}$)&(km~s$^{-1}$)       & (mas)   & (mas)       &    \\
\tableline
11587.18--11615.38  &11601.28 &\llap{$-$}0.0765 $\pm$ 0.0025 & 
0.2410  $\pm$ 0.0068       &57.4  \\
11512.57--11514.38  &11513.48 &\llap{$-$}0.0871 $\pm$ 0.0145 & 
0.2779  $\pm$ 0.0391       &10.0 \\
10834.67--10850.07  &10842.37 &\llap{$-$}0.0000 $\pm$ 0.0012 & 
0.0000  $\pm$ 0.0034       &123.4 \\
10828.33--10832.87  &10830.60 &\llap{$-$}0.0104 $\pm$ 0.0157 &
-0.0331  $\pm$ 0.0444       &  9.4 \\
10109.83--10122.44  &10116.14 & 0.1197  $\pm$ 0.0055 &
\llap{$-$}0.2908  $\pm$ 0.0155       & 26.7 \\
10099.02--10107.12  &10103.07 & 0.1297  $\pm$ 0.0116 &
\llap{$-$}0.3251 $\pm$ 0.0329       & 12.5 \\
10079.19--10092.71  &10085.95 & 0.1094  $\pm$ 0.0053 &
\llap{$-$}0.2837  $\pm$ 0.0150       & 27.6 \\
10048.55--10051.25  &10049.90 & 0.1251  $\pm$ 0.0177 &
~\llap{$-$}0.2748 $\pm$ 0.0502       &  8.2 \\
\tableline
\end{tabular}
\end{center}
\scriptsize{Notes:\\
$^a$ The range of LSR velocities (optical convention) used for re-averaging.\\
$^b$ Center of the velocity range.\\
$^c$ East–-west and north–-south position offsets from the brightest 
systemic maser. 
The rms position uncertainties reflect the fitting random errors only. \\
$^d$ The ratio of the fitted peak intensity to its rms uncertainty.}
\end{table}

\clearpage
%\begin{sidewaystable}
\begin{table}
\begin{center}
\caption{Bayesian fits for UGC~6093 and Mrk~1210 }
\label{baye}
\begin{tabular}{lcccccc}
\tableline\tableline
Galaxy    & $M_\mathrm{SMBH}^a$      & $x_0^b$    & $y_0^c$  & $PA^d$     & $i^e$   & $v_\mathrm{LSR}^f$    \\
           & ($10^7M_\odot$) & (mas)      & (mas)    & (deg)      & (deg)                    & $\mathrm{km s}^{-1}$    \\\tableline\medskip
UGC~609\rlap{3}   & 2.58$^{+0.11}_{-0.09}$            & 0.009$^{+0.008}_{-0.008}$      & 0.002$^{+0.010}_{-0.009}$    & $-20.12^{+1.50}_{-1.65}$      & 93.84$^{+4.05}_{-3.83}$                     & \llap{1}0621.17$^{+10.80}_{-12.15}$       \\
\smallskip
Mrk~1210   & 1.42$^{+0.02}_{-0.02}$             & $-$     & $-$    & \llap{$-$}$117.42^{+0.45}_{-0.45}$      & \llap{1}00.68$^{+1.13}_{-0.90}$                     & 4007.60$^{+3.60}_{-2.25}$        \\
\hline
\end{tabular}
\end{center}
Notes: the values of Mrk~1210 is for reference only.\\
$^a$ Mass of the SMBH and its 68\% fitting confidence range. \\
$^b$ East–-west offset of the SMBH from the brightest systemic maser.\\
$^c$ North–-south offset of the SMBH from the brightest systemic maser. \\
$^d$ Position angle of the disk plane (degrees east of north).\\
$^e$ Inclination angle between the disk normal and the line of sight.\\
$^f$ Optical convention LSR recession velocity.\\
%\end{sidewaystable}
\end{table}

\clearpage
\appendix 

\end{document}